\documentclass[showpacs,aps,prl,twocolumn,nofootinbib]{revtex4-2}
\usepackage{graphicx}
\usepackage[dvipsnames]{xcolor}
\usepackage{dcolumn}
\usepackage{bm}
\usepackage[hyperfootnotes=false]{hyperref}
\usepackage{lineno}
\usepackage{multirow}
\usepackage{bm}
\usepackage{diagbox}
\usepackage{makecell}
\usepackage{color}
\usepackage{graphicx}
\usepackage{xspace}

\newcommand{\gevc}{GeV/$c$\xspace}

\newcommand{\gev}{GeV\xspace}
\newcommand{\tev}{TeV\xspace}

\newcommand{\pt}{\ensuremath{p_{\rm T}}\xspace}
\newcommand{\ycm}{\ensuremath{y_{\rm CM}}\xspace}

\newcommand{\Csig}{\ensuremath{C_{\rm Sig}}\xspace}
\newcommand{\Cbg}{\ensuremath{C_{\rm Bg}}\xspace}
\newcommand{\Ctot}{\ensuremath{C_{\rm Tot}}\xspace}

\newcommand{ \la }{\langle}
\newcommand{ \ra }{\rangle}

\newcommand{\mean}[1]{\la #1 \ra}

\newcommand{\Pl}{\ensuremath{P_{\Lambda}}}

\definecolor{dgreen}{cmyk}{1.,0.,1.,0.2} 
\definecolor{orange}{cmyk}{0.,0.353,1.,0.} 

\newcommand{\sNN}{\ensuremath{\sqrt{s_{\rm NN}}}\xspace}

\newcommand{\AuAu}{Au+Au\xspace}
\newcommand{\AgAg}{Ag+Ag\xspace}

\begin{document}

\title{Measurement of global polarization of $\Lambda$ hyperons in few-GeV heavy-ion collisions}

\author{R.~Abou~Yassine$^{6,12}$, J.~Adamczewski-Musch$^{5}$, C.~Asal$^{8}$, M.~Becker$^{11}$,
A.~Belounnas$^{12}$, A.~Blanco$^{1}$, C.~Blume$^{8}$, L.~Chlad$^{13,c}$, P.~Chudoba$^{13}$,
I.~Ciepa{\l}$^{3}$, M.~Cordts$^{6}$
% ...
}

\affiliation{
(HADES collaboration) \\\mbox{$^{1}$LIP-Laborat\'{o}rio de Instrumenta\c{c}\~{a}o e F\'{\i}sica Experimental de Part\'{\i}culas , 3004-516~Coimbra, Portugal}\\
\mbox{$^{2}$AGH University of Science and Technology, Faculty of Physics and Applied Computer Science, 30-059~Kraków, Poland}\\
\mbox{$^{3}$Institute of Nuclear Physics, Polish Academy of Sciences, 31342~Krak\'{o}w, Poland}\\
\mbox{$^{4}$Smoluchowski Institute of Physics, Jagiellonian University of Cracow, 30-059~Krak\'{o}w, Poland}\\
\mbox{$^{5}$GSI Helmholtzzentrum f\"{u}r Schwerionenforschung GmbH, 64291~Darmstadt, Germany}\\
\mbox{$^{6}$Technische Universit\"{a}t Darmstadt, 64289~Darmstadt, Germany}\\
\mbox{$^{7}$Institut f\"{u}r Strahlenphysik, Helmholtz-Zentrum Dresden-Rossendorf, 01314~Dresden, Germany}\\
\mbox{$^{8}$Institut f\"{u}r Kernphysik, Goethe-Universit\"{a}t, 60438 ~Frankfurt, Germany}\\
\mbox{$^{9}$Excellence Cluster 'Origin and Structure of the Universe' , 85748~Garching, Germany}\\
\mbox{$^{10}$Physik Department E62, Technische Universit\"{a}t M\"{u}nchen, 85748~Garching, Germany}\\
\mbox{$^{11}$II.Physikalisches Institut, Justus Liebig Universit\"{a}t Giessen, 35392~Giessen, Germany}\\
\mbox{$^{12}$Laboratoire de Physique des 2 infinis Irène Joliot-Curie, Université Paris-Saclay, CNRS-IN2P3. , F-91405~Orsay , France}\\
\mbox{$^{13}$Nuclear Physics Institute, The Czech Academy of Sciences, 25068~Rez, Czech Republic}\\
\mbox{$^{14}$Uniwersytet Warszawski - Instytut Fizyki Do\'{s}wiadczalnej, 02-093~Warszawa, Poland}\\
\mbox{$^{15}$Warsaw University of Technology, 00-662~Warsaw, Poland}\\
\mbox{$^{16}$Bergische Universit\"{a}t Wuppertal, 42119~Wuppertal, Germany}\\
\\
\mbox{email:  hades-info@gsi.de}\\
\mbox{$^{a}$ also at Coimbra Polytechnic - ISEC, ~Coimbra, Portugal}\\
\mbox{$^{b}$ also at Technische Universit\"{a}t Dresden, 01062~Dresden, Germany}\\
\mbox{$^{c}$ also at Charles University, Faculty of Mathematics and Physics, 12116~Prague, Czech Republic}\\
\mbox{$^{d}$ also at Czech Technical University in Prague, 16000~Prague, Czech Republic}\\
\mbox{$^{e}$ also at University of Wroc{\l}aw, 50-204 ~Wroc{\l}aw, Poland}\\
} 

\date{\today}

\begin{abstract}
The global polarization of $\Lambda$ hyperons along the total orbital angular momentum of a relativistic heavy-ion collision is presented based on the high statistics data samples collected in \AuAu collisions at $\sNN = 2.4$~\gev and \AgAg at 2.55~\gev with the High-Acceptance Di-Electron Spectrometer (HADES) at GSI, Darmstadt.
This is the first measurement below the strangeness production threshold in nucleon-nucleon collisions.
Results are reported as a function of the collision centrality as well as a function of the hyperon's transverse momentum (\pt) and rapidity (\ycm) for the range of centrality 0--40\%.
We observe a strong centrality dependence of the polarization with an increasing signal towards peripheral collisions.
For mid-central (20 -- 40\%) collisions the polarization magnitudes are $\mean{P_\Lambda}(\%) = 6.8 \pm 1.3~\mbox{(stat.)} \pm 2.1~\mbox{(syst.)}$ for \AuAu and $\mean{P_\Lambda}(\%) = 6.2 \pm 0.4~\mbox{(stat.)} \pm 0.6~\mbox{(syst.)}$ for \AgAg, which are the largest values observed so far.
This observation thus provides a continuation of the increasing trend previously observed by STAR and contrasts expectations from recent theoretical calculations predicting a maximum in the region of collision energies about 3~GeV.
The observed polarization is of a similar magnitude as predicted by 3D-fluid-dynamics and the UrQMD plus thermal vorticity model and significantly above results from the AMPT model.
\end{abstract}
\maketitle

\mbox{}\\
\mbox{}\\
\mbox{}\\

The conversion of orbital angular momentum of a rotating rigid body into the spins of the individual particles, is rooted back to the Barnett effect, discovered already in 1915~\cite{Barnett:1915}.
Recently, this effect of mechanically induced spin polarization has been observed for electrons in liquid mercury~\cite{Takahashi:2016} and for protons in fast rotating water~\cite{Arabgol:2019}.
Here, the orbital angular momentum manifests itself in the formation of the fluid vorticity, describing the rotation of the velocity field.
A similar effect is expected in relativistic heavy-ion collisions, where orbital angular momenta of $10^3 - 10^6 \hbar$ are reached for center-of-mass energies in the nucleon-nucleon collision system of \sNN=~1~\gev--1~TeV~\cite{Becattini:2007sr}.
At collision energies \sNN~$\gtrsim$~10~\gev a nearly perfect fluid characterized by a very small shear viscosity to entropy density ratio and known as the quark-gluon plasma (QGP), is formed which might result in a spin polarization of the produced particles, commonly referred to as the global polarization ~\cite{Voloshin:2004ha,Liang:2004xn,Liang:2004ph}.

While the initial velocity fields in relativistic heavy-ion collisions result in a large anisotropy probed by collective flow, its curl, the vorticity, might cause a spin polarization of the produced particles.
Temperature gradients might also contribute significantly to the formation of vorticity similar to shear forces~\cite{Liu:2021uhn,Becattini:2021suc,Becattini:2021iol,Fu:2021pok}.
Most of the contributions will not be oriented along the orbital angular momentum direction, but might have non-zero parallel components and may depend on the location in phase-space from which the particles are emitted.
Taken all together, the vorticity field can have a complex structure~\cite{Betz:2007kg,Becattini:2017gcx,Pang:2016igs}.

The spin polarization of the produced particles can be calculated from the vorticity field, i.e. the thermal vorticity~\cite{Becattini:2015ska}.
Under the assumption of local thermal equilibrium including the spin degrees of freedom, the vorticity field is integrated over the freeze-out hypersurface to calculate the particles' polarization in the final state~\cite{Becattini:2013fla,Becattini:2020ngo}.
Since dissipative effects are neglected, this will not produce a realistic situation.
Furthermore, the separated consideration of spin potentials might be important~\cite{Bhadury:2020puc}.
For the global polarization the amount of orbital angular momentum converted to vorticity and finally to polarization of the emitted particles is relevant.
The rate of conversion is expected to grow as the collision energy is decreased and should have a maximum value at midrapidity due to the increasing role of baryon stopping~\cite{Sorge:1991pg}.
The trend can be estimated using the collision energy dependence of the directed flow~\cite{Voloshin:2017kqp}, which is also related to the velocity fields.

Experimentally, the global polarization has been observed in Au+Au collisions by the STAR Collaboration in the beam energy scan phase I (BES-I) from \sNN= 40~\gev down to \sNN= 7.7~\gev~\cite{STAR:2017ckg}.
While the QGP component can be present at all energies covered by BES-I and thus can be the main cause of the observed polarization, at lower energies the created matter will be of hadronic nature, dominated by baryons.
A recent measurement by STAR at \sNN= 3~\gev shows an increase of the global $\Lambda$ polarization to $5\,$\%~\cite{STAR:2021beb}.
For higher collision energies of \sNN= 200~\gev a significant global polarization has been measured and determined to be at the $10^{-3}$ level~\cite{Adam:2018ivw}.
At energies of a few \tev the polarization measurements by ALICE are compatible with zero and are thus consistent with an extrapolation of the energy dependence observed by STAR~\cite{ALICE:2019onw}.
In this work we investigate the global $\Lambda$ polarization at even lower collision energies with the aim to establish whether a QGP-component is mandatory for high vorticities and shed a light on the fluid-like properties of dense baryonic matter.

Several models have been used to describe the trend of the polarization~\cite{Li:2017slc,Vitiuk:2019rfv}.
The evolution above this energy in the intermediate range up to RHIC energies is very uncertain.
Different scenarios are predicted by the available models, thus stressing the importance to collect new experimental data in this energy region.
A study of thermal and kinematic vorticity within the UrQMD model~\cite{Bleicher:1999xi} resulted in a maximum value of the global polarization at around \sNN = 3-5~\gev, depending on the collision centrality~\cite{Deng:2020ygd}.
Calculations using a 3D-fluid-dynamics model based on a thermodynamic approach and taking the non-equilibrium effects at the early stages of the collision into account by two counterstreaming baryon-rich fluids, predict the maximum at \sNN= 2.4~\gev~\cite{Ivanov:2020udj}.
Another study using the AMPT model~\cite{Lin:2004en} locates the maximum around \sNN = 7.7 \gev~\cite{Guo:2021udq}.
Hence, global polarization measurements at collision energies \sNN$< 7.7$~\gev will provide important constraints on the model calculations and will further promote our understanding of the nature of QCD matter created at collision energies of a few \gev.
Studies should also be made as a function of centrality, as the global polarization is expected to approach zero in central collisions and increase linearly towards peripheral events together with the orbital angular momentum increase.
A weak dependence with respect to the transverse momentum is expected, as particles with \pt=~0 are still subject to the vorticity of the surrounding matter.

In this letter, differential results are presented as functions of transverse momentum (\pt) center-of-mass rapidity (\ycm) and collision centrality for the global $\Lambda$ polarization in \AgAg collisions at \sNN= 2.55~\gev and \AuAu collisions at \sNN= 2.42~\gev measured with HADES.
Data are placed in the context of results from other experiments and compared to modern model calculations.
Information on the flow pattern at lower energies is provided by detailed measurements with HADES for protons, deuterons and tritons~\cite{HADES:2020lob}.

The measurement of the global $\Lambda$ polarization is performed using the self-analyzing weak decay $\Lambda \rightarrow p + \pi^-$, in which the daughter proton is preferentially emitted along the $\Lambda$ spin direction in the $\Lambda$ rest frame.
The orientation of the orbital angular momentum is estimated via the reaction plane (RP), spanned by the impact parameter of the colliding nuclei and the beam direction. 
The RP is approximated by the event plane (EP), which can be reconstructed from the deflection of the projectile spectators in the direction of the impact parameter~\cite{Poskanzer:1998yz}.

\begin{figure}[th]
\includegraphics[width=1\linewidth]{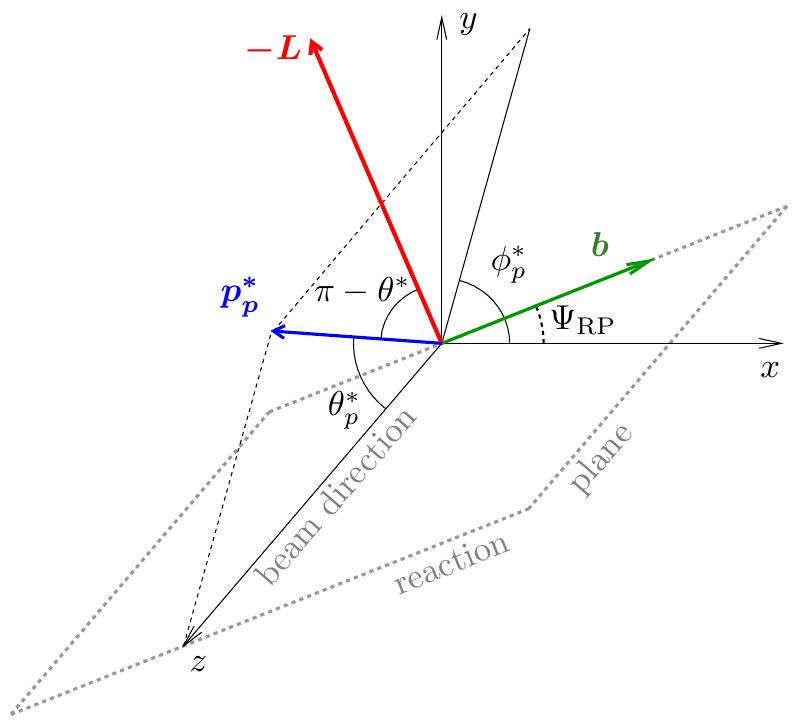}
\caption{
\label{notaionsPlot}
Diagram showing the notations for the different angles adopted in this paper.
The laboratory frame is defined by the $x$, $y$, and $z$ (beam direction) axes.
{\boldmath $p^*_p$} is the decay proton 3-momentum in the $\Lambda$ rest frame.
The reaction plane is spanned by the impact parameter {\boldmath $b$} and the beam direction.
The normal to the reaction plane defines the direction of the system's orbital momentum {\boldmath $L$}. Reversal of the orbital momentum, {\boldmath $L$}$\to$ -{\boldmath $L$}, corresponds to changing the reaction plane angle by $\Psi_{\rm RP}\to \Psi_{\rm RP}+\pi$.
}
\end{figure}

The global polarization observable is defined by~\cite{STAR:2007ccu}:
\begin{equation}
\Pl = \frac{8}{\pi \alpha_\Lambda} \frac{\langle \sin(\Psi_{\rm EP} - \phi_{\rm p}^*)\rangle}{ R_{\rm EP}}.
\label{Eq:1}
\end{equation}
Here $\alpha_\Lambda = 0.732 \pm 0.014$~\cite{ParticleDataGroup:2020ssz} is the $\Lambda$ decay parameter, $\Psi_{\rm EP}$ the event plane angle, $\phi_{\rm p}^*$ the azimuthal angle of the proton in the $\Lambda$ rest frame, $R_{\rm EP}$ the resolution of the event plane angle and the brackets $\langle . \rangle$ denote the average over all produced $\Lambda$ hyperons.
A schematic drawing of the geometrical measures is shown in Fig.~\ref{notaionsPlot}.
As the polarization is measured in \%, we will give absolute numbers and omit the "\%" at the end to avoid confusion, whenever effects of corrections or systematic uncertainties are discussed in the following.

HADES is a fixed-target experiment located at GSI, Darmstadt~\cite{HADES:2009aat}.
Its acceptance is well suited for midrapidity measurements of $\Lambda$ hyperons.
It has a large polar angle ($\theta$) coverage of $18^\circ$ to $85^\circ$ and a full coverage in the azimuthal angle.
A 15-fold segmented target is used to increase the interaction rate and to allow for the use of thin target foils which guarantee a sufficient vertex position resolution along the beam axis to identify weakly decaying particles.
The Multi-Wire Drift Chambers (MDCs), two in front and two behind the toroidal magnetic field, allow for tracking and momentum determination.
A measurement of the time-of-flight is performed with the Resistive Plate Chamber (RPC), which covers the polar angle region of $18^\circ < \theta < 45^\circ$ and the Time-Of-Flight detector (TOF) covering $44^\circ < \theta < 85^\circ$.
Combining the information of these two subsystems with the measurement of the start time of the collision provided by the mono-crystalline diamond based (START) detector allows for particle identification.
The EP is reconstructed using the spectator projectiles measured by the Forward Wall (FW) detector, which is placed $6.8\,$m behind the target and covers $0.34^\circ < \theta < 7.4^\circ$.

The global polarization results presented here are based on $11 \times 10^{9}$ \AgAg collisions recorded at \sNN= 2.55~\gev by HADES in 2019, and $2 \times 10^{9}$ \AuAu collisions recorded at \sNN= 2.42~\gev in 2012.
The energy of the \AuAu collisions is below the nucleon-nucleon production threshold for the $\Lambda$ hyperon of 2.55~\gev, and the energy is exactly at threshold for the \AgAg collisions.
The collision centrality is determined using the hit multiplicity of the TOF and RPC by using calculations with the Glauber model~\cite{HADES:2017def}.
Events within $0-40\,$\% centrality are used in this analysis.
The event selection has been further optimized to reject reactions with the carbon holder and pile-up events.

The $\Lambda$ hyperons are reconstructed via their weak decay $\Lambda \rightarrow p + \pi^-$.
The decay products are identified using loose cuts on the reconstructed mass determined from the time-of-flight together with the measured momenta.
Subsequently, the decay products are combined and a set of parameters describing the decay topology are calculated and used in a multi-variate analysis (MVA) in order to distinguish true $\Lambda$ hyperons from the combinatorial background~\cite{HADES:2018noy,Kornas:2020qzi}. 
A selection efficiency of $60\,$\% ($87\,$\%) is achieved at a background suppression rate of $97\,$\% ($92\,$\%) for the \AuAu (\AgAg) data, respectively.

\begin{figure*}
    \begin{minipage}{0.49\textwidth}
    \centering
    \includegraphics[width=1\linewidth]{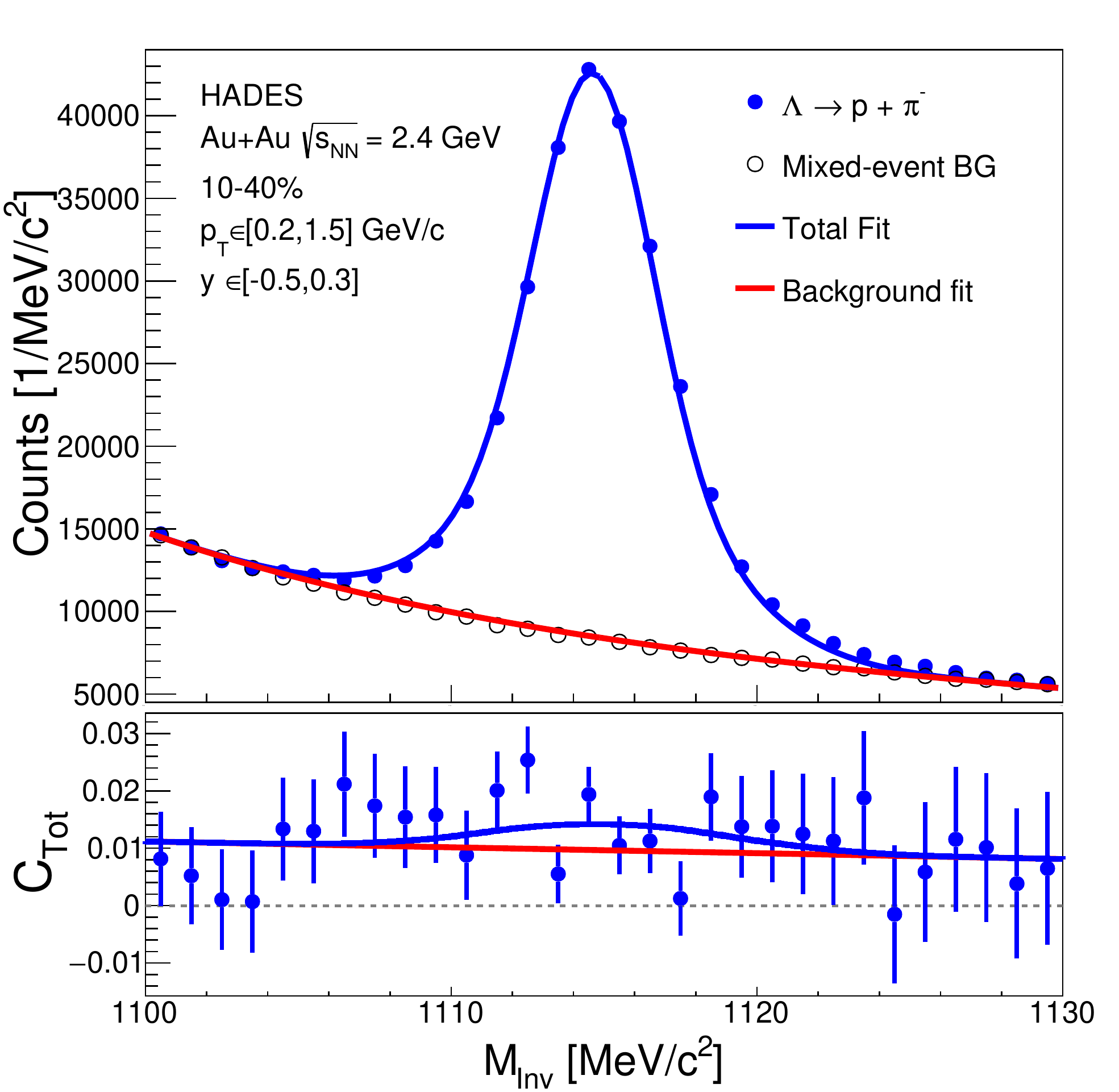}
    \end{minipage}
    \hfill
    \begin{minipage}{0.49\textwidth}
    \centering
    \includegraphics[width=1\linewidth]{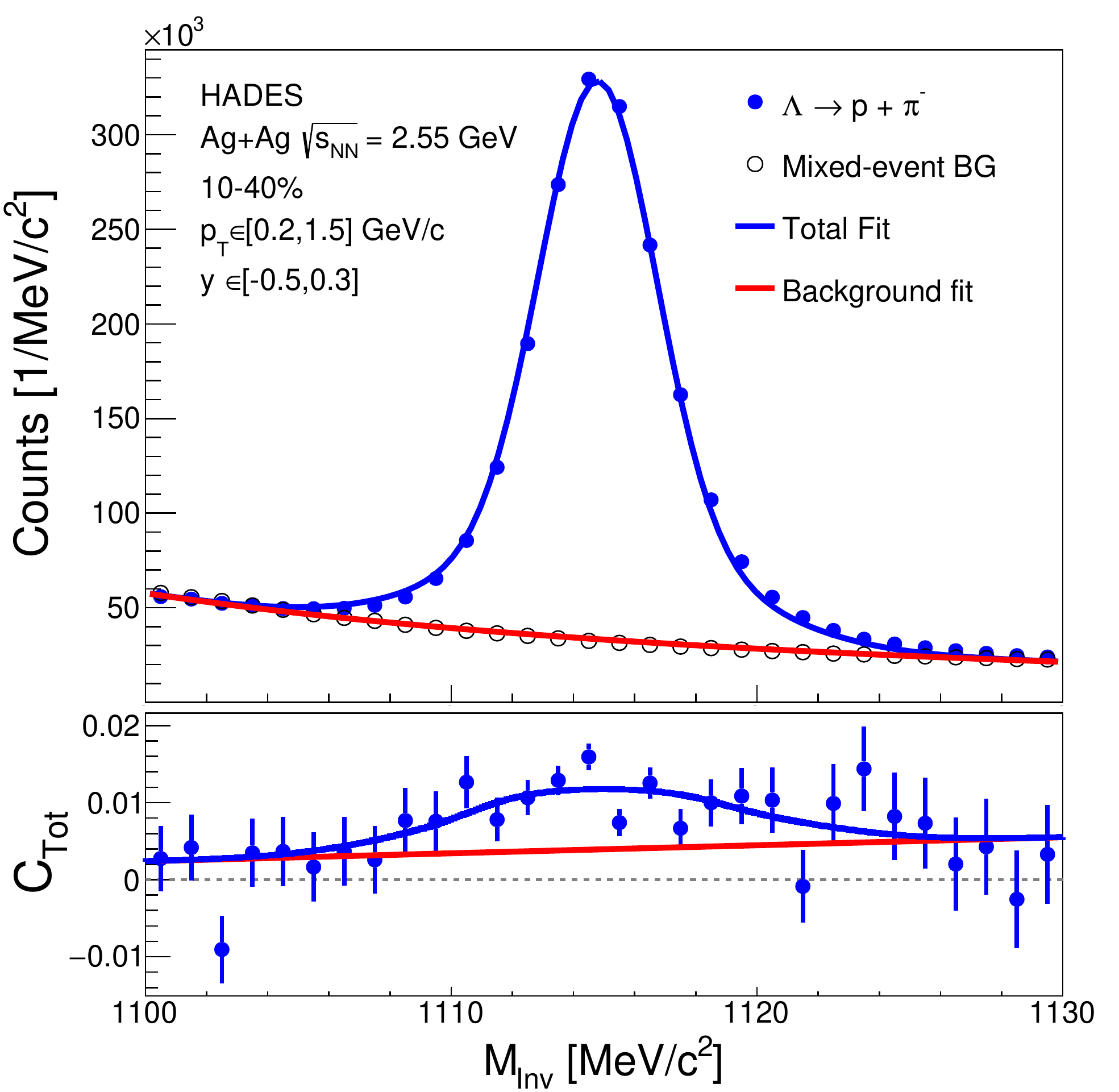}
    \end{minipage}
    \caption{Illustration of the performance of the $\Lambda$ hyperon reconstruction and polarization extraction of signal events using data taken with HADES.
    Only statistical uncertainties are shown.
    The top panels show the pion-proton invariant-mass distribution for selected candidates with $0.2 < \pt < 1.5$ \gevc and $-0.5 < \ycm < 0.3$ in \AuAu (left) and \AgAg (right) collisions at \sNN = 2.4 and 2.55~\gev with the centrality 10--40\%.
    The bottom panels show the invariant mass dependence of the $\Ctot(M_{\rm Inv}) = \langle \sin(\Psi_{\rm EP} - \phi_{\rm p}^*)\rangle$ from Eq.~\ref{Eq:1} for the same $\Lambda$ candidates selection as depicted in the top panel.}
    \label{Fig:1}
\end{figure*}

The overall reconstruction efficiency depends on the transverse momentum and rapidity and ranges from $3$ to $12\,$\% in \AuAu and $5$ to $15\,$\% in \AgAg~\cite{Kornas:2022cbl}.
Figure~\ref{Fig:1} (top panels) shows an example of the invariant mass distribution of $\Lambda$ hyperons after efficiency correction.
The combinatorial background is described by a Landau function fit to the mixed-event data.
The signal peak has been fitted with a double Gaussian function, and a $\pm 2\sigma$ range is used to determine the significance and signal-to-background ratio.
In total, $0.19\,$M ($1.5\,$M) $\Lambda$ hyperons with a signal-to-background ratio of $2.3$ ($5.0$) are reconstructed in the $10-40\,$\% centrality range for \AuAu (\AgAg).
Besides the difference in the overall number of events, the multiplicity of $\Lambda$ hyperons per event is lower in \AuAu collisions since the beam energy is below the threshold for $\Lambda$ production~\cite{HADES:2018noy}.
Due to the larger system size, the combinatorial background is increased with respect to the \AgAg data.
The polarization of the $\Lambda$ hyperons, being proportional to $\Csig = \langle \sin(\Psi_{\rm EP} - \phi_{\rm p}^*)\rangle$ in Eq.~(1), was extracted using the invariant-mass fit method~\cite{Borghini:2004ra,STAR:2013ayu}.
The signal contribution \Csig to the total correlation $\Ctot(M_{\rm Inv})=\langle \sin(\Psi_{\rm EP} - \phi_{\rm p}^*)\rangle_{\rm Tot}$ is separated from the background contribution $\Cbg(M_{\rm Inv})=\langle \sin(\Psi_{\rm EP} - \phi_{\rm p}^*)\rangle_{\rm Bg}$ as a function of the invariant mass according to:
\begin{equation}
\Ctot(M_{\rm Inv}) = f_{\rm Sig}(M_{\rm Inv})\Csig + f_{\rm Bg}(M_{\rm Inv})\Cbg(M_{\rm Inv}).
\end{equation}
Here, $f_{\rm Sig}$ and $f_{\rm Bg}$ are the corresponding signal and background fractions as extracted from the fit in Fig.~\ref{Fig:1}~(top).
The background correlation is observed to be non-zero, i.e. $\Cbg(M_{\rm Inv}) \neq 0$.
\Cbg has been investigated using Monte Carlo (MC) simulations to study correlations between polarization, anisotropic flow and detector acceptance \cite{Kornas:2022cbl}.
The invariant mass dependence of $\Cbg$ has been reproduced and traced back to a mismatching of decay daughters from polarized $\Lambda$ hyperons combined to primary particles. 
Thereby, the dominant contribution has been identified to be combinations of $\pi^-$ from polarized $\Lambda$ hyperons with protons not emerging from $\Lambda$ decays due to the large amount of primary protons in the collision.
To describe the background correlation in the experimental data, a linear dependence $\Cbg(M_{\rm Inv}) = \alpha + \beta \, M_{\rm Inv}$ is assumed as shown in Fig.~\ref{Fig:1} (bottom). 

$\Ctot(M_{\rm Inv})$ has been weighted by the inverse of the event plane resolution, efficiency and radial distance asymmetry (RDA)~\cite{Kornas:2022cbl}.
The event plane resolution is determined by randomly grouping FW hits in two equally sized sub-events $A$ and $B$ and then calculating $R_{\rm EP} = \langle \text{cos}(\Psi_{\rm A} - \Psi_{\rm B}) \rangle$ as described in~\cite{Poskanzer:1998yz}.
$R_{\rm EP}$ has been calculated for centrality  classes with $5\,$\% width, and for mid-central collisions the values range from $0.82$--$0.88$ ($0.66$--$0.75$) for \AuAu (\AgAg)~\cite{HADES:2020lob}.
The efficiency correction is applied as a function of rapidity and transverse momentum.
Only relative variations of the efficiency for the different selected phase-space regions are taken into account.
The RDA correction is related to the cross product of the proton track direction and the beam axis.
Depending on its sign, a strong variation of the $\Lambda$ polarization and yield is observed.
The latter is used to derive the correction applied to the polarization averaged over the full RDA range.
The same dependence is reproduced in MC simulations and is related to a strong correlation of the RDA to the directed flow of the particles.
Therefore, the data have been corrected for the RDA resulting in an enhancement of the extracted polarization signal of $+0.2$ in both data sets.

For the evaluation of the systematic uncertainties the Barlow criterion~\cite{Barlow:2002yb} was used to determine whether a given variation of $P_{\rm V}$ relative to the measured value $P_{\rm M}$ with corresponding statistical uncertainties $\sigma_{\rm V}$ and $\sigma_{\rm M}$ is statistically significant: 
$B_{\rm crit} = |P_{\rm M} - P_{\rm V}| ~\left / \sqrt{|\sigma_{\rm M}^2 - \sigma_{\rm V}^2|}\right.$.
Variations with $B_{\rm crit} > 1$ were added in quadrature to obtain the total systematic uncertainty.
A detailed description of all systematic sources considered including a summary table can be found in~\cite{Kornas:2022cbl}.
Among the sources of systematic uncertainties are those originating from the selection of the $\Lambda$ hyperons, with the most prominant one being the selection on the distance of closest approach (DCA) of the proton track to the event vertex, which contributed $\pm 0.72$ ($\pm 0.28$) to the overall systematic errors for \AuAu (\AgAg) collisions.
In \AuAu collisions, the variation of the MVA response and the effect of the efficiency correction cause systematic uncertainties of similar magnitude, $\pm 0.59$ and $\pm 0.66$ respectively.
Both contributions were found to be negligible in \AgAg.
A second method, the $\Delta \phi$-extraction method~\cite{Poskanzer:1998yz}, has been used to evaluate systematic uncertainties originating from the method applied.
No significant variation beyond statistical fluctuations in comparison to the invariant-mass fit method has been observed.
This is also valid for variations of the RDA correction procedure which do not pass the Barlow criterion.
In the systematic uncertainty, a variation of the decay parameter by $\pm 0.014$~\cite{ParticleDataGroup:2020ssz} and of the event plane resolution by $3\,$\% ($5\,$\%) relative variation for \AuAu (\AgAg) collisions are included. 
The latter is based on the variations of $R_{\rm EP}$ using sub-divisions of the FW hits according to the different cell sizes and comparing the results between different combinations of the subevents.

For the differential analysis in \AgAg, most of the systematic variations are propagated from the integrated result in order to reduce statistical fluctuations due to the smaller data sets for the individual bins.
Only those sources expected to depend on phase-space or centrality respectively, are re-evaluated bin-by-bin~\cite{Kornas:2022cbl}, as for example the uncertainty on the correction for the event plane resolution ranges from 15\% (0--10\% centrality) to 3\% (30--40\% centrality) in relative numbers.
Other sources are related to the background determination which can be very different depending on phase-space and centrality.
These are: the modeling of the background shape in the invariant-mass fit method, the RDA and efficiency correction as well as the $\Delta \phi$-extraction method.

To quantify the interplay between polarization and directed flow, the analysis is also performed as a function of $\phi_\Lambda - \phi_{\rm p}^*$.
From this distribution a Fourier decomposition can be performed, where the constant term allows to extract the overall polarization~\Pl.
Even though a significant contribution from the directed flow is observed, it is only reflected in the relative modulations of \Pl~as a function of $\phi_\Lambda - \phi_{\rm p}^*$ but not in the integrated result.

Due to the lower charged particle multiplicity in \AgAg collisions the peripheral events are contaminated with Ag+C events of similar multiplicity originating from collisions of beam ions with the carbon target holder.
These collisions are in general not symmetric with respect to the beamline and therefore covered by the RDA correction. 
The effect of the RDA correction is $\pm 0.2$ of the extracted polarization signal which is within the assigned total systematic uncertainty.

\begin{figure}
    \centering
    \includegraphics[width=1\linewidth]{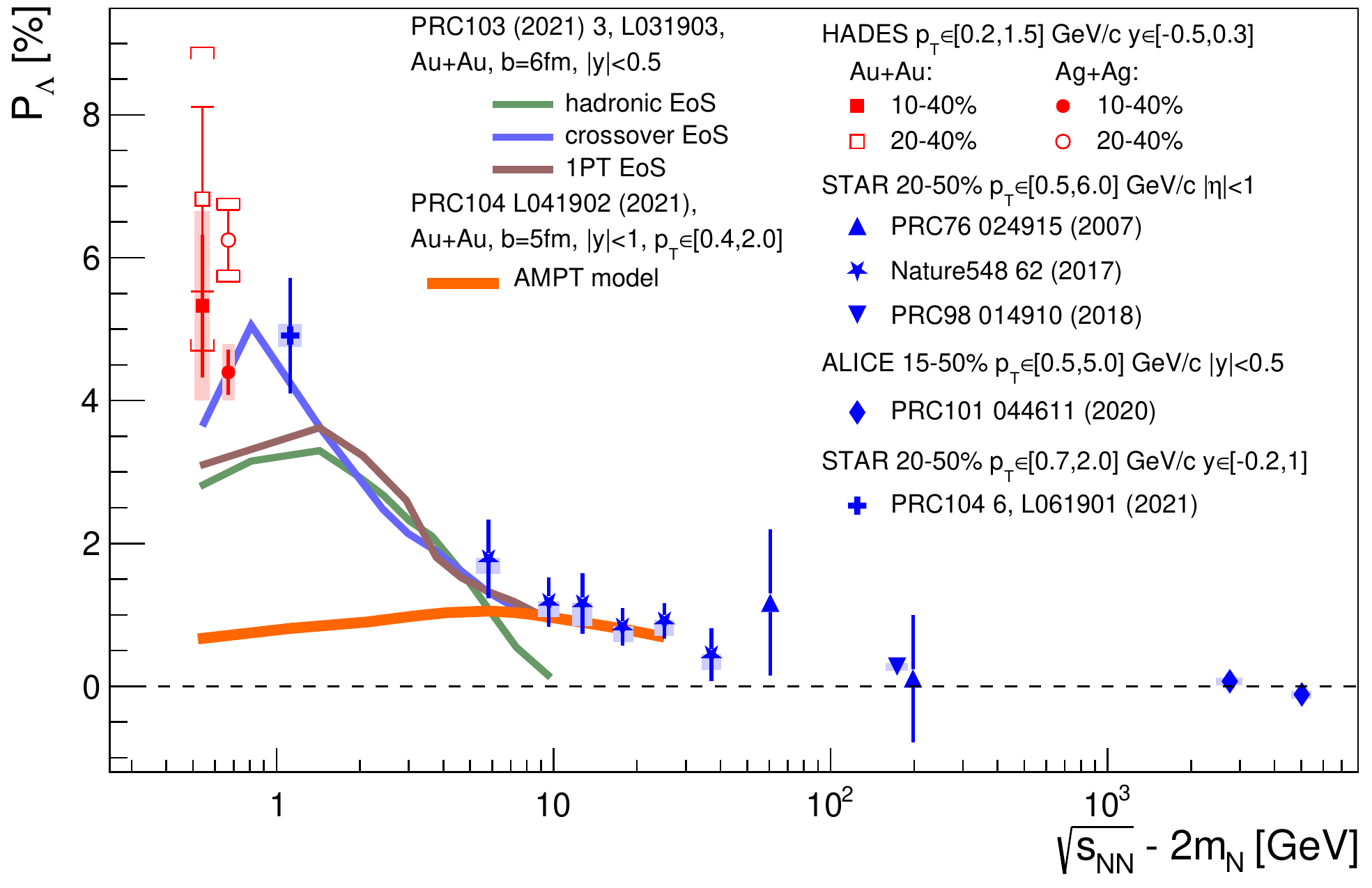}
    \caption{Global polarization of $\Lambda$ hyperons as a function of the center-of-mass energy above $2m_{\rm N}$, where $m_{\rm N}$ is the nucleon mass.
    Statistical uncertainties are indicated by the error bars attached to the data points and the systematic uncertainties are represented by the boxes.
    All results are scaled to the currently accepted value of the decay parameter $\alpha_\Lambda = 0.732$~\cite{ParticleDataGroup:2020ssz}.
    The model calculations based on 3D-fluid-dynamics~\cite{Ivanov:2020udj} are shown as solid lines (green, blue, brown) for three different EoSs. 
    The red solid line represents the prediction by the AMPT model, assuming a direct connection between the polarization vector and the thermal vorticity in thermal equilibrium \cite{Guo:2021udq}.}
    \label{Fig:2}
\end{figure}

Figure~\ref{Fig:2} shows the collision energy dependence of \Pl.
The HADES data are shown for $0.2 < \pt < 1.5$ \gevc and $-0.5 < \ycm < 0.3$ in the 10-40\% centrality range.
The data from the RHIC BES-I program and fixed-target run by the STAR collaboration and the measurements by ALICE at LHC are shown for comparison.
The ALICE measurements are scaled with the latest PDG value of the hyperon decay constant~\cite{ALICE:2019onw}.
To avoid premature conclusions on the location of the maximum global polarization, the HADES data are shown for 20-40\% centrality too.
A clear enhancement with respect to the 10-40\% results is observed indicating the strong centrality dependence of the global $\Lambda$ polarization.
This is also important for the comparison to other measurements, expecially to the STAR 3~\gev result which is shown for 20-50\% centrality.
The 20-40\% HADES data indicate a continuation of the increasing global $\Lambda$ polarization towards lower collision energies.

The data are compared to different model calculations, performed for the \AuAu system and averaged over impact parameter to match 10-40\% in collision centrality. 
Strikingly, our data confirm that AMPT model calculations drastically underestimate the global $\Lambda$ polarization below \sNN $\leq 10$~\gev.
Such a discrepancy could point to the presence of a significant effect related to the frictional interaction of the participants with the spectators of the collision, which increases the angular momentum transfer but is not included in the calculations~\cite{Guo:2021udq}.
In contrast, calculations based on 3D-fluid dynamics~\cite{Ivanov:2020udj} are able to reproduce the observed magnitude and the increasing trend towards SIS18 energies.
Three different assumptions on the equation-of-state (EoS), namely crossover, first-order phase transition, and a purely hadronic, have been used for these calculations but with the current precision of the HADES measurements these scenarios cannot be distinguished.

\begin{figure}
    \centering
    \includegraphics[width=1\linewidth]{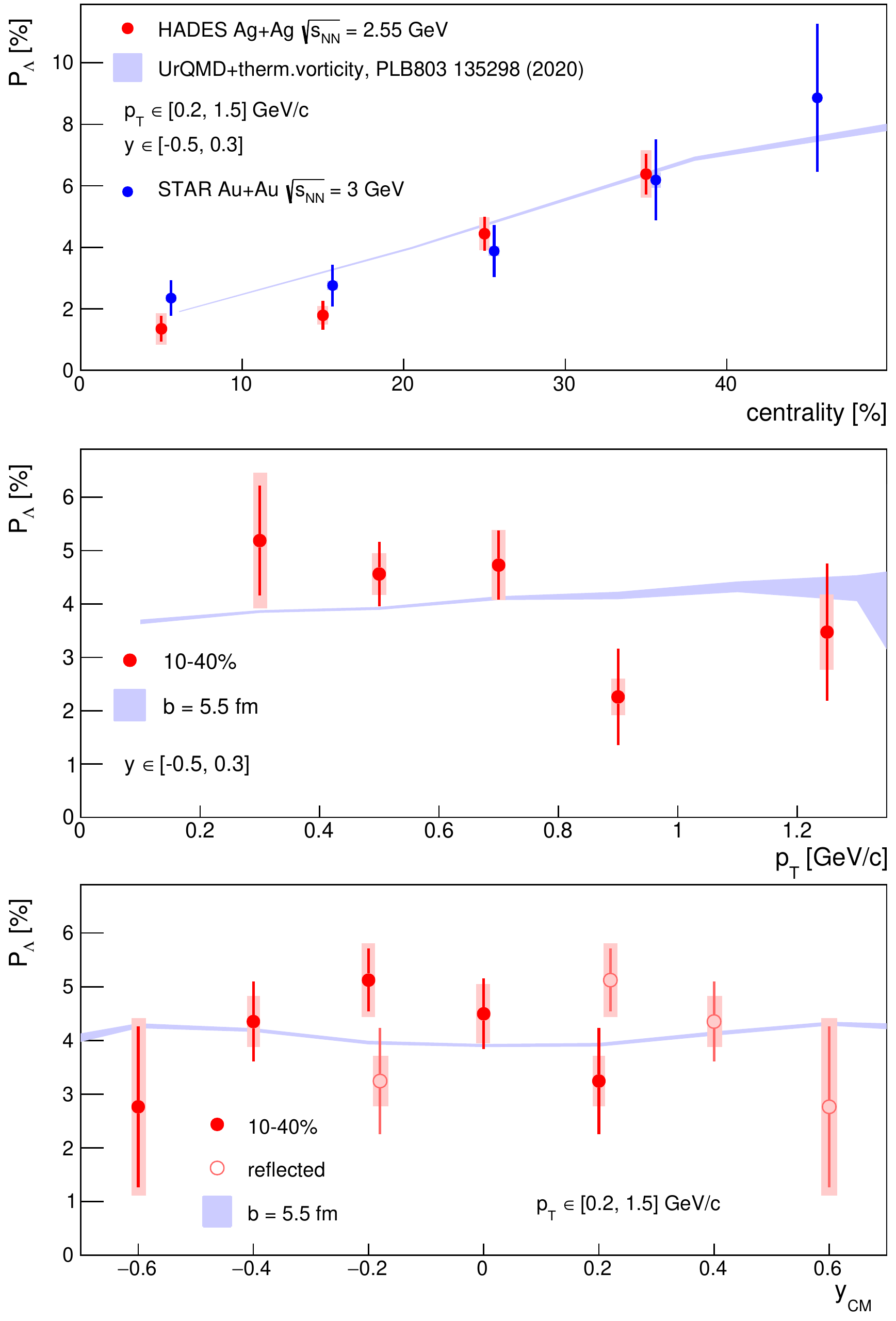}
    \caption{Global polarization of $\Lambda$ hyperons as a function of (top) collision centrality, (middle) transverse momentum and (bottom) rapidity for \AgAg collisions.
    Results for the centrality dependence are shown for $0.2 < \pt < 1.5$ \gevc and $-0.5 < \ycm < 0.3$ while results as a function of \pt and \ycm are shown for the 10--40\% centrality class.
    The reflected points are shifted for better visibility.
    The centrality dependence measured by STAR in Au+Au at \sNN=~3~\gev is shown for comparison~\cite{STAR:2021beb}.
    The results are compared to UrQMD model calculations (light blue band) based on the thermal vorticity approach~\cite{Vitiuk:2019rfv}.}
    \label{Fig:3}
\end{figure}

Figure~\ref{Fig:3} shows the results for \Pl~as a function of collision centrality, transverse momentum and rapidity in \AgAg collisions.
To illustrate the expected symmetry in \ycm, the data points mirrored around \ycm=~0 are shown as open symbols.
A strong centrality dependence is observed with the largest signal for mid-central (30--40\%) collisions and vanishing \Pl~for central collisions.
This is consistent with the evolution of the orbital angular momentum, which in a simplified approach is expected to grow linearly with the impact parameter \cite{Becattini:2007sr}.
The centrality dependence in \AgAg at \sNN= 2.55~\gev is very consistent with the STAR 3~\gev measurement in Au+Au, except for the most central collisions.
Also, note that the selected phase-space regions differ between the different experiments as summarized in the legend of Fig.~\ref{Fig:2}.
The HADES data do not show a strong variation of \Pl with \pt or \ycm within acceptance, although a slight fall off towards higher \pt and an increase towards midrapidity cannot be excluded.
In general, all results can be reproduced by calculations based on the UrQMD model plus thermal vorticity calculations~\cite{Vitiuk:2019rfv}.

In summary, the global polarization of $\Lambda$ hyperons is measured in \AuAu collisions at $\sNN = 2.42$~\gev and \AgAg at 2.55~\gev by the HADES experiment at GSI, Darmstadt.
The integrated hyperon global polarization $\mean{P_\Lambda}(\%)$ is found to be significant and amounts to $5.3 \pm 1.0~\mbox{(stat.)} \pm 1.3~\mbox{(syst.)}$ for \AuAu and $\mean{P_\Lambda}(\%) = 4.4 \pm 0.3~\mbox{(stat.)} \pm 0.4~\mbox{(syst.)}$ for \AgAg collisions in the centrality range 10--40\%.
In terms of the collision energy dependence, the HADES results are in agreement with 3D-fluid-dynamical calculations and disfavor the predictions based on the AMPT model which systematically underestimate the global $\Lambda$ polarization at the lower collision energies.
The \AgAg data are also reported as a function of the collision centrality, 0--40\%, as a function of transverse momentum, $0.2 < \pt < 1.5\,$GeV/$c$, and rapidity, $-0.7 < \ycm < 0.3$.
The results are compatible with calculations based on the UrQMD plus thermal vorticity model, although the model slightly overpredicts the measured data.
Nevertheless, the agreement is remarkable as dissipative and non-equilibrium effects have not yet been taken into account.

These HADES results on the global polarization open a new window for the study of vortical structure of baryon dominated matter created in heavy-ion collisions at the energies of a few GeV per nucleon.
Remarkably, the global polarization is found to be highest around the strangeness production threshold of \sNN~=~2.55~GeV, but in turn must vanish around \sNN$\sim 2m_N \approx 1.9\,$GeV.
This poses puzzling questions on the origin of the polarization mechanism, as the global $\Lambda$ polarization increases continuously from the deconfined matter (QGP) to the baryon-dominated hadronic matter at lower collision energies.
These results are essential for the extraction of details on equilibration time, evolution dynamics, and the equation-of-state of QCD matter and extend the set of existing measurements available towards lower energies.

\textbf{Acknowledgments.} The HADES collaboration acknowledge O.~Vitiuk and J.~Liao for providing calculations and for elucidating discussions. SIP JUC Cracow, Cracow (Poland), 2017/26/M/ST2/00600; TU Darmstadt, Darmstadt (Germany), VH-NG-823, DFG GRK 2128, DFG CRC-TR 211, BMBF:05P18RDFC1; Goethe-University, Frankfurt (Germany) and TU Darmstadt, Darmstadt (Germany), DFG GRK 2128, DFG CRC-TR 211, BMBF:05P18RDFC1, HFHF, ELEMENTS:500/10.006, VH-NG-823, GSI F\&E, ExtreMe Matter Institute EMMI at GSI Darmstadt; JLU Giessen, Giessen (Germany), BMBF:05P12RGGHM; IJCLab Orsay, Orsay (France), CNRS/IN2P3; NPI CAS, Rez, Rez (Czech Republic), MSMT LTT17003, MSMT LM2018112, MSMT OP VVV CZ.02.1.01/0.0/0.0/18\_046/0016066.
\\
\\
The following colleagues from Russian institutes did contribute to the results presented in this publication but are not listed as authors following the decision of the HADES Collaboration Board on March 23, 2022:
A.~Belyaev, D.~Borisenko, O.~Fateev, O.~Golosov, M.~Golubeva, F.~Guber, A.~Ierusalimov, A.~Ivashkin, N.~Karpushkin, V.~Khomyakov, V.~Ladygin, A.~Lebedev, M.~Mamaev, S.~Morozov, O.~Petukhov, A.~Reshetin, S.~Reznikov, A.~Shabanov, A.~Taranenko, A.~Zhilin, and A.~Zinchenko. 

\nocite{*}

\bibliography{references}

\end{document}